%% file: FSQ-15-001_temp.tex
\pdfoutput=1

\documentclass[11pt,twoside,a4paper,cmspaper,final,collab]{cms-tdr}
\def\svnVersion{310121}\def\cmsCernNo{2015-180}\def\cmsCernDate{\today}\def\cmsMessage{Submitted to Physics Letters B}

\begin{document}\cmsNoteHeader{FSQ-15-001}

\hyphenation{had-ron-i-za-tion}
\hyphenation{cal-or-i-me-ter}
\hyphenation{de-vices}
\RCS$Revision: 306592 $
\RCS$HeadURL: svn+ssh://alverson@svn.cern.ch/reps/tdr2/papers/FSQ-15-001/trunk/FSQ-15-001.tex $
\RCS$Id: FSQ-15-001.tex 306592 2015-10-23 14:31:50Z sikler $
\newlength\cmsFigWidth
\ifthenelse{\boolean{cms@external}}{\setlength\cmsFigWidth{0.85\columnwidth}}{\setlength\cmsFigWidth{0.4\textwidth}}
\ifthenelse{\boolean{cms@external}}{\providecommand{\cmsLeft}{top\xspace}}{\providecommand{\cmsLeft}{left\xspace}}
\ifthenelse{\boolean{cms@external}}{\providecommand{\cmsRight}{bottom\xspace}}{\providecommand{\cmsRight}{right\xspace}}

\newcommand {\dnchdeta} {\ensuremath{\rd{}N_{\mathrm{ch}}/\rd\eta}\xspace}
\newcommand{\Peight}{{\textsc{pythia8}}\xspace}
\newcommand{\EPOS}{{\textsc{epos}\,LHC}\xspace}
\newcommand{\meta}{2}
\newcommand{\finaldat}{\ensuremath{5.49 \pm 0.01\stat \pm 0.17\syst}\xspace}

\cmsNoteHeader{FSQ-15-001}

\title{Pseudorapidity distribution of charged hadrons in proton-proton
collisions at $\sqrt{s} = 13$\TeV}

\date{\today}

\abstract{
The pseudorapidity distribution of charged hadrons in pp collisions at
$\sqrt{s} =13$\TeV is measured using a data sample obtained with the CMS
detector, operated at zero magnetic field, at the CERN LHC. The yield of
primary charged long-lived hadrons produced in inelastic pp collisions is
determined in the central region of the CMS pixel detector ($\abs{\eta}<2$) using
both hit pairs and reconstructed tracks.  For central pseudorapidities
($\abs{\eta}<0.5$), the charged-hadron multiplicity density is $\dnchdeta|_{\abs{\eta}
< 0.5} = \finaldat$, a value obtained by combining the two methods.
The result is compared to predictions from Monte Carlo event generators and to
similar measurements made at lower collision energies.}

\hypersetup{%
pdfauthor={CMS Collaboration},%
pdftitle={Pseudorapidity distribution of charged hadrons in proton-proton collisions at sqrt(s) = 13 TeV},%
pdfsubject={CMS},%
pdfkeywords={CMS, physics, proton-proton, 13 TeV, hadrons, spectra}}

\maketitle

\section{Introduction}

The yields of charged hadrons are among the most basic physical observables in
high-energy particle collisions and provide an essential first step in
exploring the physics of a new energy regime. Studies of such yields have a
long history in high-energy particle and nuclear physics, as well as in cosmic
ray physics.
At collider energies, the inclusive production of charged hadrons is driven by
a combination of perturbative and nonperturbative quantum chromodynamics (QCD)
phenomena, such as saturation of parton densities, multiparton interactions,
parton hadronization, and soft diffractive scattering.

The yields of primary charged hadrons are commonly studied using their
multiplicity as a function of pseudorapidity, $\dnchdeta$.
Of particular interest for understanding the physics of hadron production is
the dependence of $\dnchdeta$ on the collision energy, which reflects the
relative roles of soft- and hard-scattering contributions. Soft interactions,
which are modeled phenomenologically, give rise to a significant fraction of
the produced particles. Contributions from hard-scattering processes increase
with increasing collision energies. Measurements are necessary to tune the
modeling of these contributions in Monte Carlo (MC) event generators, and as
reference data to study nuclear effects in proton-nucleus and nucleus-nucleus
collisions.
A good understanding of inclusive hadron production is also important to
control the pileup backgrounds, from overlapping proton-proton collisions in a
given bunch crossing, that affect all physics analyses at the LHC.

In this Letter, measurements of $\dnchdeta$ in the range $\abs{\eta} < \meta$ are
reported for inelastic proton-proton (pp) collisions delivered by the CERN LHC
at a center-of-mass energy of 13\TeV in June 2015.
The analysis is based on 11.5\,million events recorded at zero magnetic field
during a special low-intensity beam configuration with 0.2--5\% proton-proton
interaction probability per bunch crossing. This special run was prepared by
steering the beams such that their transverse separation was $\pm 3\sigma$ at
the nominal CMS interaction point, where $\sigma$ denotes the standard
deviation of the Gaussian beam profile.
Following earlier analyses at $\sqrt{s} = 0.9\TeV$,
2.36\TeV~\cite{Khachatryan:2010xs}, 7\TeV~\cite{Khachatryan:2010us}, and
8\TeV~\cite{Chatrchyan:2014qka}, $N_{\rm ch}$ is defined to include decay
products of particles with decay length $c\tau < 1\unit{cm}$, where $\tau$ is
the lifetime of the particle and $c$ is the velocity of light. Products of
secondary interactions are excluded, and contributions from prompt leptons are
removed.

The data are compared to \Peight v208~\cite{Sjostrand:2006za,Sjostrand:2007gs}
(with the CMS underlying event tunes~\cite{GEN-14-001}: CUETP8S1 and CUETP8M1,
using different parton densities), and to \EPOS~\cite{Werner:2005jf} (LHC
tune~\cite{Pierog:2013ria}). Both MC event generators reproduce well the main
characteristics of the experimental data measured in hadronic collisions at
lower energies, and provide predictions for the $\sqrt{s}$-dependence of hadron
production observables using different implementations of the dominant
underlying phenomena (multiparton interactions, parton saturation, diffractive
scattering)~\cite{d'Enterria:2011kw,Khachatryan:2015fia}.

\section{CMS detector}

The central feature of the CMS apparatus is a superconducting solenoid of
6\unit{m} internal diameter. Within the magnet volume are the silicon pixel and
strip tracker, the crystal electromagnetic calorimeter, and the brass and
scintillator hadron calorimeter.
The tracker measures charged particles within $\abs{\eta} < 2.4$. It has 1\,440
silicon pixel and 15\,148 silicon strip detector modules, arranged in 14
tracking layers.
The barrel region of the CMS pixel detector consists of three layers very
close to the beam line. They are located at average radii of 4.3\unit{cm}
(layer~1), 7.2\unit{cm} (layer~2), and 11.0\unit{cm} (layer~3), and provide
excellent position resolution with their $150\mum{\times} 100\mum$ pixels. The
data sample used for this analysis was obtained while the magnet was off. Under
these conditions, charged particles follow approximately straight trajectories,
perturbed mostly by multiple Coulomb scattering. The pixel hits alone are
sufficient to reconstruct vertices and tracks with high precision and purity.
The beam pickup for timing (BPTX) devices were used to trigger the detector
readout. They are located around the beam pipe at a distance of 175\unit{m}
from the interaction point (IP) on either side, and are designed to provide
precise information on the LHC bunch structure and the timing of the incoming
beams.
A detailed description of the CMS detector can be found in
Ref.~\cite{Chatrchyan:2008aa}.

\section{Event selection}

Inelastic collision events are selected as follows: (i) online, a coincidence
of signals from both BPTX devices is required, indicating the presence of both
proton bunches crossing the IP; (ii) offline, at least one reconstructed
interaction vertex is required, according to the tracklet or track vertex
reconstruction methods described in Sections 4.1 and 4.2, respectively.
While only the reconstructed collision vertex with the highest multiplicity
(primary vertex) is used in the tracklet analysis, all reconstructed vertices
in a given bunch crossing are processed for the track analysis.

A study of noncolliding bunches shows that the above requirements are
sufficient to reject all backgrounds not originating from pp collisions. The
probability to select events in the presence of a single (noncolliding) beam is
about $5 \times 10^{-5}$ per bunch crossing, while the same probability is
smaller than $3 \times 10^{-6}$ per bunch crossing in the absence of any beam.
Consequently, the contribution of background events from beam,
beam halo, and cosmic ray sources to the observed yields is negligible.

The results are corrected to correspond to a sample of inelastic collisions.
The corrections from the detector-level offline event selection to the
hadron-level event definitions are derived from MC simulations with
\Peight v208 (tune CUETP8M1) and \EPOS\ (tune LHC), which cover a wide
range of possible model ingredients.
The detector response is simulated with \GEANTfour~\cite{Agostinelli2003250}
and processed through the same event reconstruction chain as collision data.
The MC simulations are produced with the location and shape of the interaction
region as extracted from data.

\section{Data analysis}

The analysis of the recorded data is performed with two reconstruction
techniques based on hits of charged particles detected by the CMS pixel
detector. While both start by searching for hit pairs in different layers at
similar azimuthal angles $\phi$, the tracklet method (using hit pairs) performs
background subtraction based on control samples in data, while the track method
(using hit triplets) minimizes background contributions by requiring an
additional hit in the detector. This results in a slightly narrower accessible
$\eta$ range for the track method as compared to the tracklet method.
Since various factors, such as detector alignment, material, detector response,
and dependence on MC event generators, influence the two techniques somewhat
differently, their final combination into a single measurement provides a more
robust result.

\subsection{Tracklets}

Tracklets are pairs of hits in two layers of the silicon pixel detector. For a
tracklet that is consistent with a charged hadron that originates at the
primary vertex, the difference in pseudorapidity ($\Delta\eta$) and the
azimuthal angle ($\Delta\phi$) between the
two hits that make up the tracklet is small. The correlation between the hits
in two silicon layers is analyzed to determine the charged hadron $\dnchdeta$.
This method is capable of measuring and correcting for the combinatorial
background and it is sensitive to hadrons with transverse momentum $\pt$ as low
as 0.04\GeVc.

\begin{figure}[thb]
 \centering
  \includegraphics[width=0.49\textwidth]{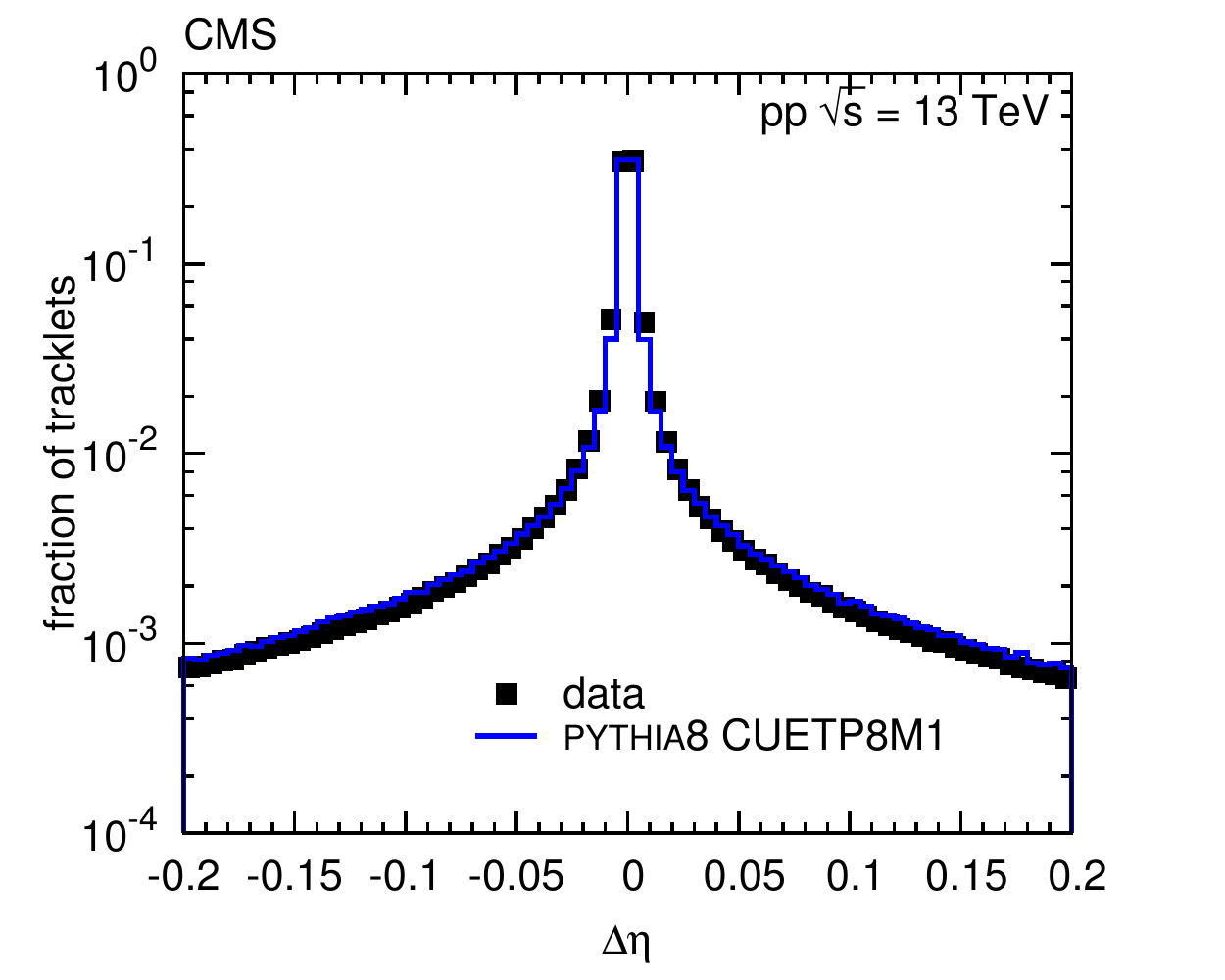}
  \includegraphics[width=0.49\textwidth]{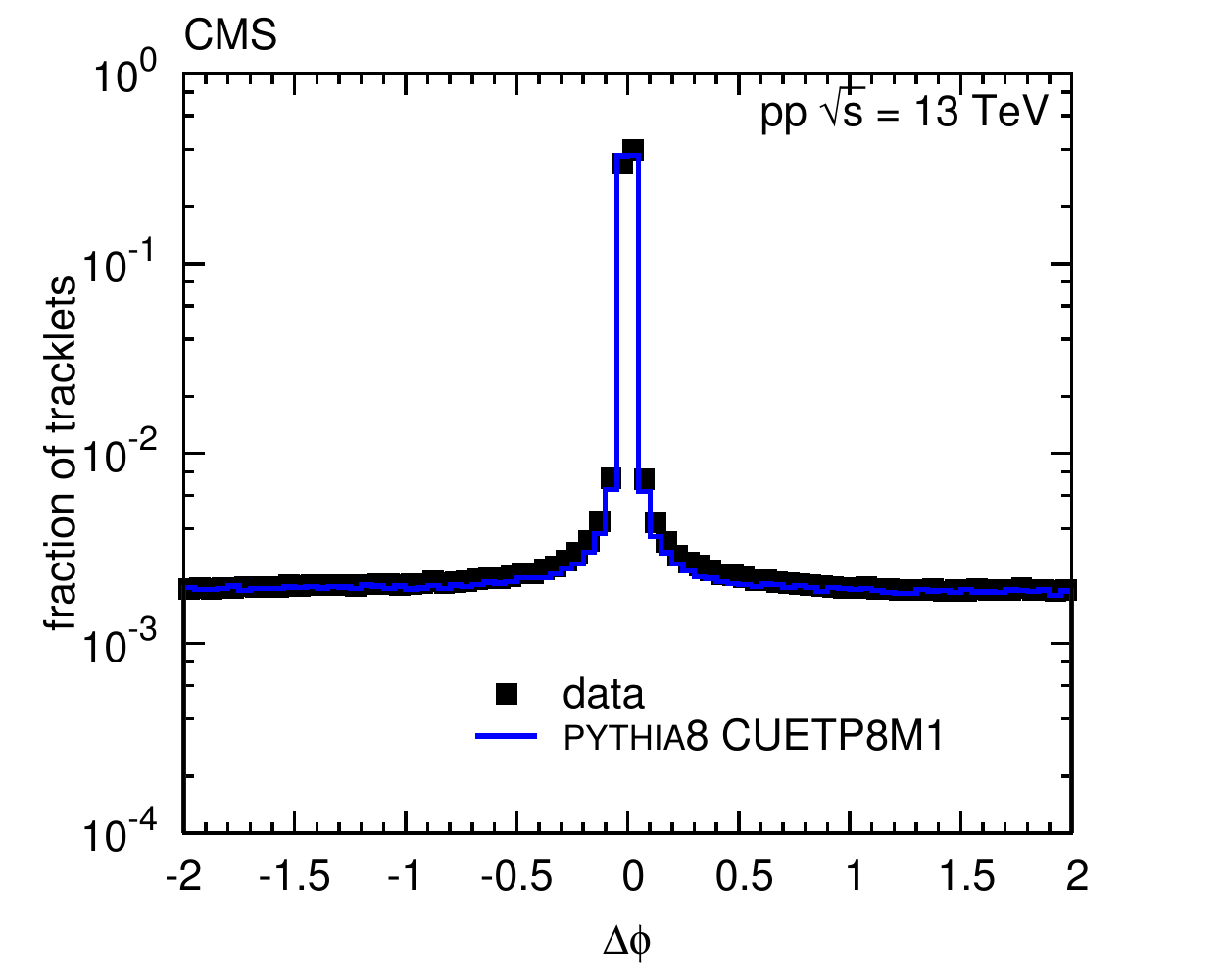}
 \caption{The $\Delta\eta$ (\cmsLeft) and $\Delta\phi$ (\cmsRight) distributions of hit
pairs on tracklets in the data (squares) and from MC simulation (histogram).}
 \label{fig:tracklet}
\end{figure}

A tracklet-based vertex finder, as described in Ref.~\cite{Chatrchyan:2011pb},
is used for primary vertex reconstruction. In the first step, a hit from
layer~1 is selected and a matching hit from layer~2 is sought. If the
$\abs{\Delta\phi}$ of the hits is smaller than 0.05, the $z$ position with respect
to the counterclockwise-beam direction of the hits is extrapolated linearly and
projected onto the beam axis. This procedure is repeated for every hit in
layer~1, and the calculated $z$ positions are saved as vertex candidates. The
primary vertex is determined in a second step. If the magnitude of the
difference between the $z$ positions of any two vertex candidates is smaller
than 0.12\unit{cm}, they are combined into a vertex cluster. The vertex cluster
with the highest number of associated vertex candidates is selected as the
primary vertex, and the final vertex $z$ position $z_v$ is given by the average
$z$ position of the associated vertex candidates. The typical resolution of the
primary vertex $z$ position is 0.02--0.1\unit{cm}, depending on the number of
pixel hits. The vertex reconstruction efficiency is high even for
low-multiplicity events with few pixel hits, with around 80\% efficiency for
events with 3 to 4 hits in layer~1, and 100\% efficiency for events with more
than 8 hits in layer~1.

The tracklet reconstruction is a separate procedure from the vertex
reconstruction. There is no requirement on the $\Delta\phi$ of the hits.
Instead, hits with the smallest $\abs{\Delta\eta}$  are paired first to form
tracklets (Fig.~\ref{fig:tracklet}, \cmsLeft), and no hit may be used more than
once. In contrast to the track analysis, all hits are accepted for the
tracklet reconstruction, such that the analysis is relatively insensitive to
the cluster charge simulation. Tracklets corresponding to charged hadrons that
originate at the primary vertex display a sharp peak at $\Delta\phi=0$, while
the background tracklets from uncorrelated pixel hits have a wide $\Delta\phi$
spectrum. Figure~\ref{fig:tracklet} (\cmsRight) shows the combined data of primary
charged hadrons (sharp peak) and tracklets from uncorrelated pixel hits (wide
tails). To suppress the combinatorial background, only tracklets with
$\abs{\Delta\eta}<0.1$ and $\abs{\Delta\phi}<1$ are selected. Since the combinatorial
background is flat in $\Delta\phi$, a sideband region $1<\abs{\Delta\phi}<2$ can be
defined and used to estimate its magnitude, which is then subtracted from the
signal region $\abs{\Delta\phi}<1$ to obtain the uncorrected $\dnchdeta$. Typical
values of the estimated background fraction in the signal region increase with
$\abs{\eta}$ from 6\% to 25\%. The $\eta$ range for the tracklet method is limited
to $\abs{\eta}<\meta$ to avoid a large acceptance correction.

Contributions from secondary particles, as well as the tracklet acceptance and
reconstruction efficiency, have to be accounted for in order to determine the
charged hadron $\dnchdeta$ distribution. These correction factors are
calculated as a function of the primary vertex $z$ position, pseudorapidity,
and tracklet multiplicity, using the \Peight CUETP8M1 tune as a reference.

To account for the difference between the actual pixel detector geometry and
that used in the simulation, an additional correction is applied as a function
of $\eta$ and the primary vertex position. This correction is obtained by
taking the ratio between data and simulation of the geometrical distribution of
tracklets in ($\eta$, $z_v$) bins.
The size of this correction ranges from 0 to 5\%.
The largest correction factors are associated with the presence of inactive
tracker modules in the data.

\subsection{Tracks}

The analysis only uses clusters whose width in the $z$ direction is compatible
with that of a charged particle originating from the nominal collision point.
The difference between the measured and the geometrically predicted widths is
required to be smaller than 5 pixels.

Track finding involves identifying pixel hit triplets that fall on a straight
line. All possible hit pairs are taken, the first hit from layer~1, and the
second one from layer~2. If the difference between their azimuthal angles
$\phi_1$ and $\phi_2$ is smaller than a certain value, ${\Delta\phi_{1,2}} <
\alpha$, they are kept for the next step.
For each hit pair, hits from layer~3 that have a small azimuthal difference
with respect to the hit in layer~2 ($\abs{\Delta\phi_{2,3}} < \alpha$) are
collected. In addition, the vector joining the hits in the first and second
layers and the vector joining the hits in the second and third layers are
required to have polar angles that differ by, at most, $\alpha$
($\abs{\Delta\theta_{12,23}} < \alpha$). If multiple hits in the third layer
satisfy the conditions above, the one with the smallest $\Delta\phi_{2,3}^2 +
\Delta\theta_{12,23}^2$ value is selected.
The comparison of the $\Delta\phi$ distributions from signal and background
shows that a value of $\alpha = 0.02$ gives the best signal significance.

The coordinates of the hits of the resulting triplets are used to perform a
straight-line fit with parameters $(z_0, \eta, \phi)$ using Newton's method for
minimization, where $z_0$ is the $z$ coordinate of the point of closest
approach to the beam axis. The transverse impact parameter $d_0$ with respect
to the beam axis is fixed to 0 for this fit.
The average distance $\overline{d}$ from the fitted line to the hits is used in
the determination of the vertex position to estimate the uncertainty in $z$, as
$\sigma_z = \overline{d}/\sin\theta$.
A track is accepted if the straight-line fit to the hits gives a vertex
position $\abs{z_0} < 20\unit{cm}$.

The acceptance of the track method is $\abs{\eta} < 1.8$, slightly reduced compared
to the tracklet method, since all three pixel barrel layers are used.
According to the samples of pp events generated with \Peight CUETP8M1 and
\EPOS, the tracking efficiency is flat at 80\% in the region $\abs{\eta} < 1.6$
owing to losses at low $\pt$, and increases to 85\% for $\pt > 0.2\GeVc$. For
$\abs{\eta} > 1.6$, the efficiency falls and is on average around 50\%. The
wide $\pt$ coverage down to 0.05\GeVc ensures a robust performance and fairly
insensitive behavior to variations in the $\pt$ spectrum. The rate of duplicate
tracks is below the percent level, while the fraction of misidentified tracks
is in the range of 2--6\%, rising with higher $\abs{\eta}$ values. The fraction of
reconstructed nonprimary particles is 2--3\%.

\begin{figure}[t!bhp]
\centering
\includegraphics[width=0.49\textwidth]{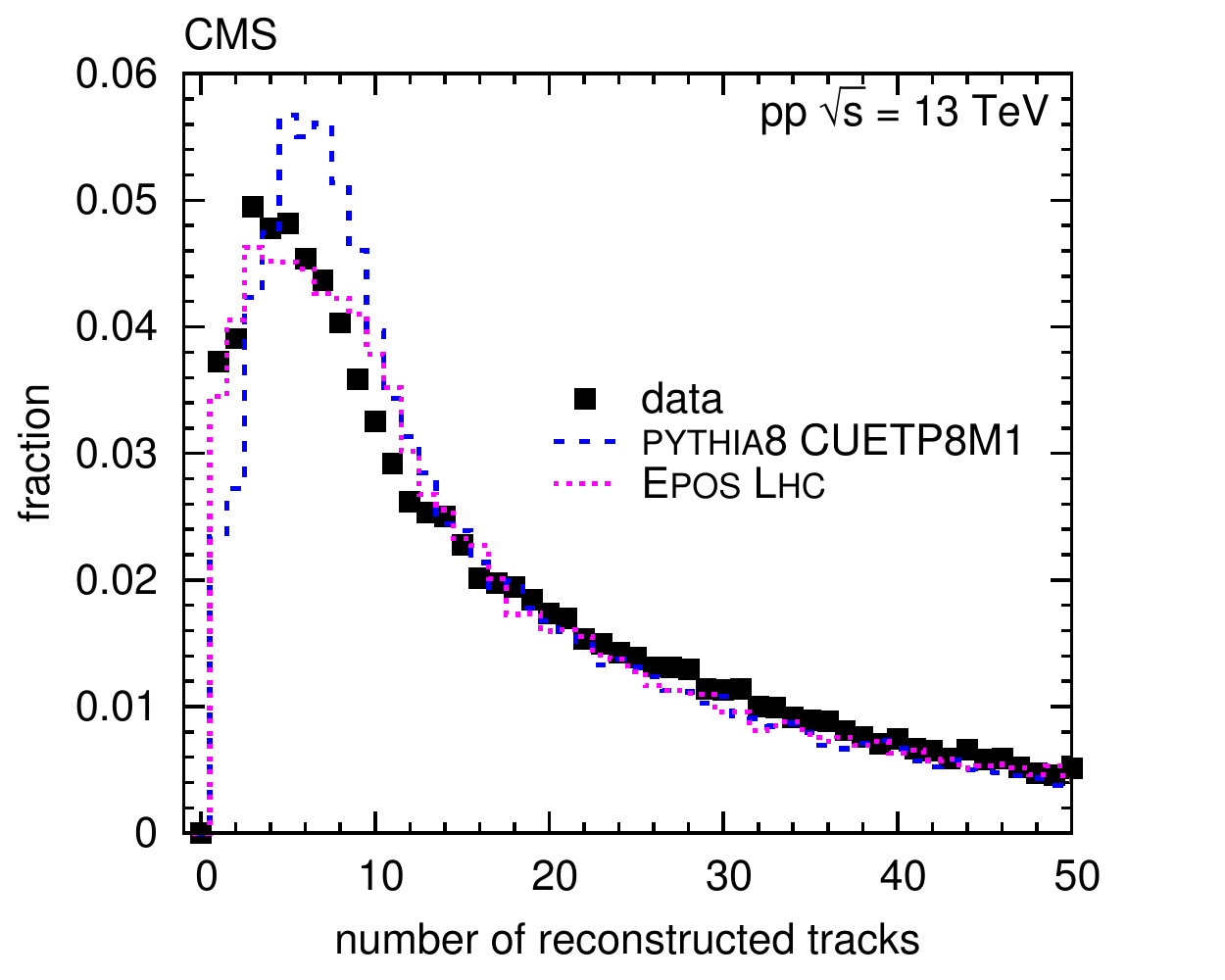}
\includegraphics[width=0.49\textwidth]{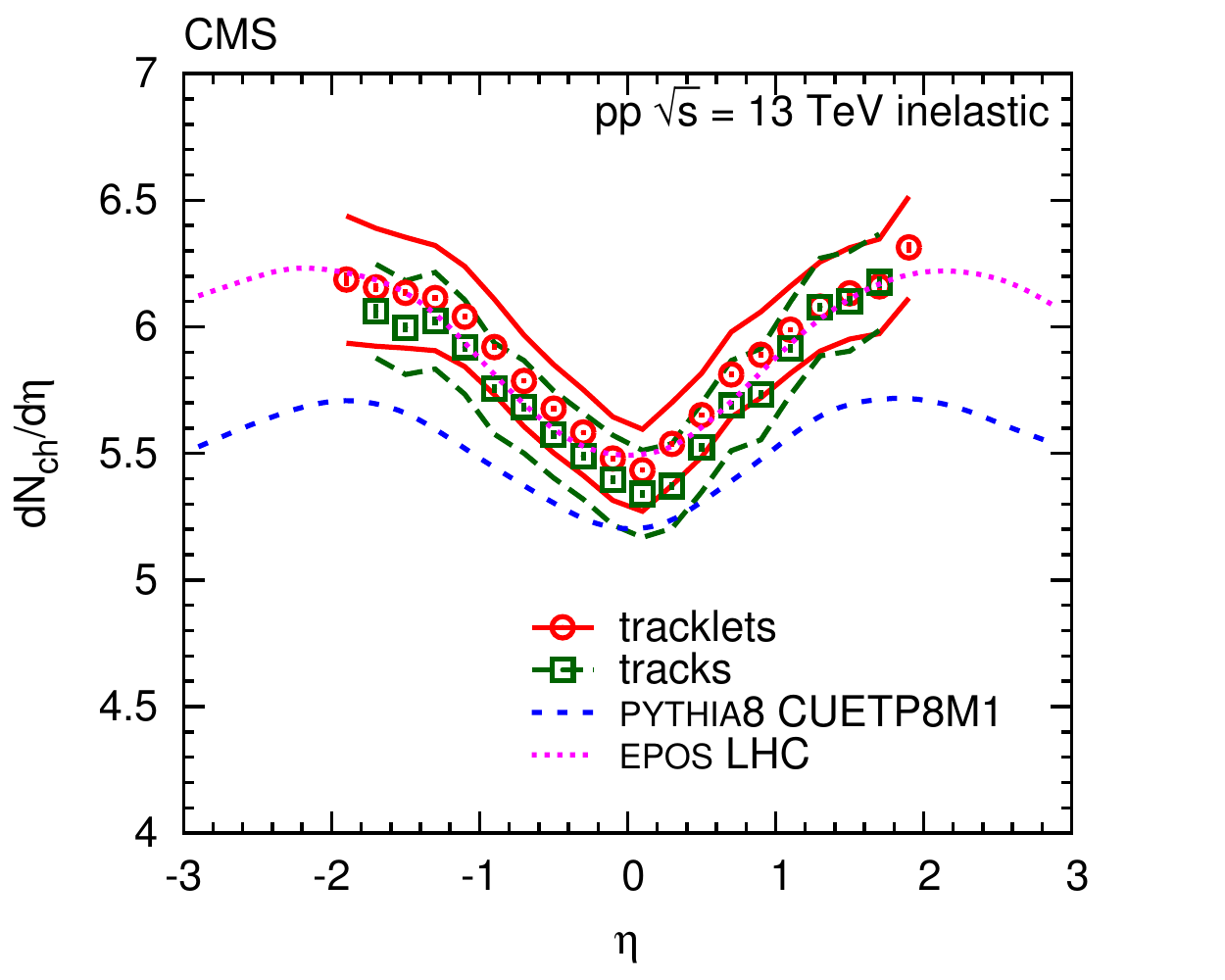}
\caption{(\cmsLeft) Distributions of the number of reconstructed tracks per event
from data and from simulations, for inelastic pp collisions.
(\cmsRight) Distributions of the pseudorapidity density (\dnchdeta) of charged
hadrons in the region $\abs{\eta} < \meta$ for inelastic collision events,
reconstructed using tracklets (open circles), and tracks (open squares),
compared to \Peight CUETP8M1 and \EPOS predictions, shown as dashed and dotted
curves, respectively.
The solid and long-dashed lines encompassing the data points indicate the
systematic uncertainties of the two reconstruction methods.}
\label{fig:tracking_performance}
\end{figure}

An agglomerative vertex reconstruction~\cite{Sikler:2009nx} is performed using
the fitted $(z_0, \sigma_z)$ values from the reconstructed tracks.
The clustering of tracks into vertices ends when the smallest
$\sigma_z$-normalized distance between vertex candidates is larger than 50.
Duplicate tracks are removed based on the angle enclosed by their direction
vectors. If this angle is smaller than 0.01, the one with a larger average
distance $\overline{d}$ is removed.
Vertices with at least three tracks are kept, unless there is only one vertex
found in the event. In that case, there is no minimum for the number of tracks.
Only tracks associated with a primary vertex are used in the physics analysis.

The hits of the final tracks are refitted by allowing the impact parameter
$d_0$ to vary. The resulting set of parameters is used in a three-dimensional
vertex fit, which is also used to determine the location and shape of the
interaction region.
As Fig.~\ref{fig:tracking_performance} (\cmsLeft) shows, the distribution of the
number of reconstructed tracks in data is in fair agreement with the MC event
generator predictions in particular at the higher multiplicities.
The correction from track to charged particle multiplicity depends on the
fraction of detected events as well as on the distribution of high multiplicity
events (not shown in the figure).

Owing to the excellent vertex resolution ($\approx$250\mum), the $\eta$
resolution is also very good. Thus, potential migrations between neighboring
bins are very small, and the final generator-dependent corrections are applied
bin-by-bin. The measured yields in each $\eta$ bin are multiplied by the ratio
of the number of simulated primary charged hadrons to the number of the
reconstructed tracks.

The performance of vertexing in simulation and in data was compared by
determining the probability of reconstructing $k$ vertices in case of $n = 1$,
2, and 3 simultaneous pp collisions.
The extracted $p(k|n)$ conditional probabilities are used to estimate the
average inelastic interaction probability $\mu$ per bunch crossing by
assuming a Poissonian distribution for $n$.
While \Peight CUETP8M1 gives $\mu = 0.0525 \pm 0.0003\stat$, \EPOS leads to
$0.0545 \pm 0.0002\stat$. Taking their average as the central value and the
difference as the systematic uncertainty, we estimate $\mu = 0.0535 \pm
0.0013\syst$. The product of the number of analyzed bunch crossings and $\mu$
gives the number of inelastic collision events, to be used for the calculation
of the final $\dnchdeta$ values.

\subsection{Systematic uncertainties}

The systematic uncertainties in the final physics result come from several
sources.
According to the two MC event generators used, the two reconstruction
techniques detect about 86--89\% (for the tracklet method) and 86--90\% (for
the track method) of the inelastic pp events. The uncertainty in the number of
unseen, mostly single-diffractive, collisions contributes about 3\% to the
total systematic uncertainty.

The systematic uncertainties in the tracklet method take into account the
effects of the amount of noise in the pixel hits (1--3\%), detector
misalignment, pixel hit reconstruction efficiency, and cluster splitting (all
below 1\%). The tracklet-to-hadron correction shows a 2--3\% dependence on the
choice of the MC event generator. The total systematic uncertainty of the
tracklet method is in the range 3--4\%.

For the track method, requiring three hits on a straight line greatly reduces
the effects of pixel noise and cluster splitting. In addition,
hits with cluster shapes that are not compatible with their positions are
excluded from the analysis, and tracks close in $(\eta,\phi)$ are cleaned as
described in Section 4.2.
The loss of any of the three hits because of pixel reconstruction inefficiency
would lead to the loss of the triplet, resulting in a systematic uncertainty
of 1.8\% from the tracking efficiency, independent of the choice of the MC
event generator.
The effect of detector misalignment on the angular
distributions is several orders of magnitude smaller than the values of the
selection criteria, and is therefore neglected.
The sensitivity to the vertexing efficiency is included in the 2--3\%
systematic uncertainty associated with the estimate of $\mu$, as discussed in
Section 4.2.
Uncertainties on track-level corrections are estimated from the differences
obtained using the two MC generators, and are at the level of 2--3\%. The total
systematic uncertainty of the track method is in the range of 3--4\%.

The track analysis was also performed using only matched hits of double-sided
modules in the strip tracker. The result is compatible with that from the
pixel-only analysis within the systematic uncertainties. Because of the larger
corrections and the more restricted $\abs{\eta}$ range, the result with the strip
tracker is not included in the final combination.

Finally, we also used the method of counting reconstructed pixel hits, as
described in Ref.~\cite{Khachatryan:2010us}, which is subject to
different background and systematic effects, and the measurements are within
the systematic uncertainties of the results reported in this Letter.

\section{Results}

For the tracklet analysis, all recorded bunch crossings are used (about
170\,000 collision events), while for the track analysis only 1 million of them
are used (about 55\,000 collision events).
After corrections, the agreement between the tracklet and track $\dnchdeta$
results is better than 2\% at central pseudorapidity and better than 3\% at
forward pseudorapidities, as shown in Fig.~\ref{fig:tracking_performance}
(\cmsRight). Hence, averaging their $\dnchdeta$ values is justified. Since the
systematic uncertainties dominate and are mostly correlated between the two
analyses, the simple mean of the central values and of systematic uncertainties
are taken.
The fraction of primary charged leptons is $\approx$1\% of the total
long-lived charged particles produced, and the correction from detector-level
tracklets and tracks is done for charged hadrons only.

\begin{figure}[t!bhp]
 \centering
  \includegraphics[width=0.49\textwidth]{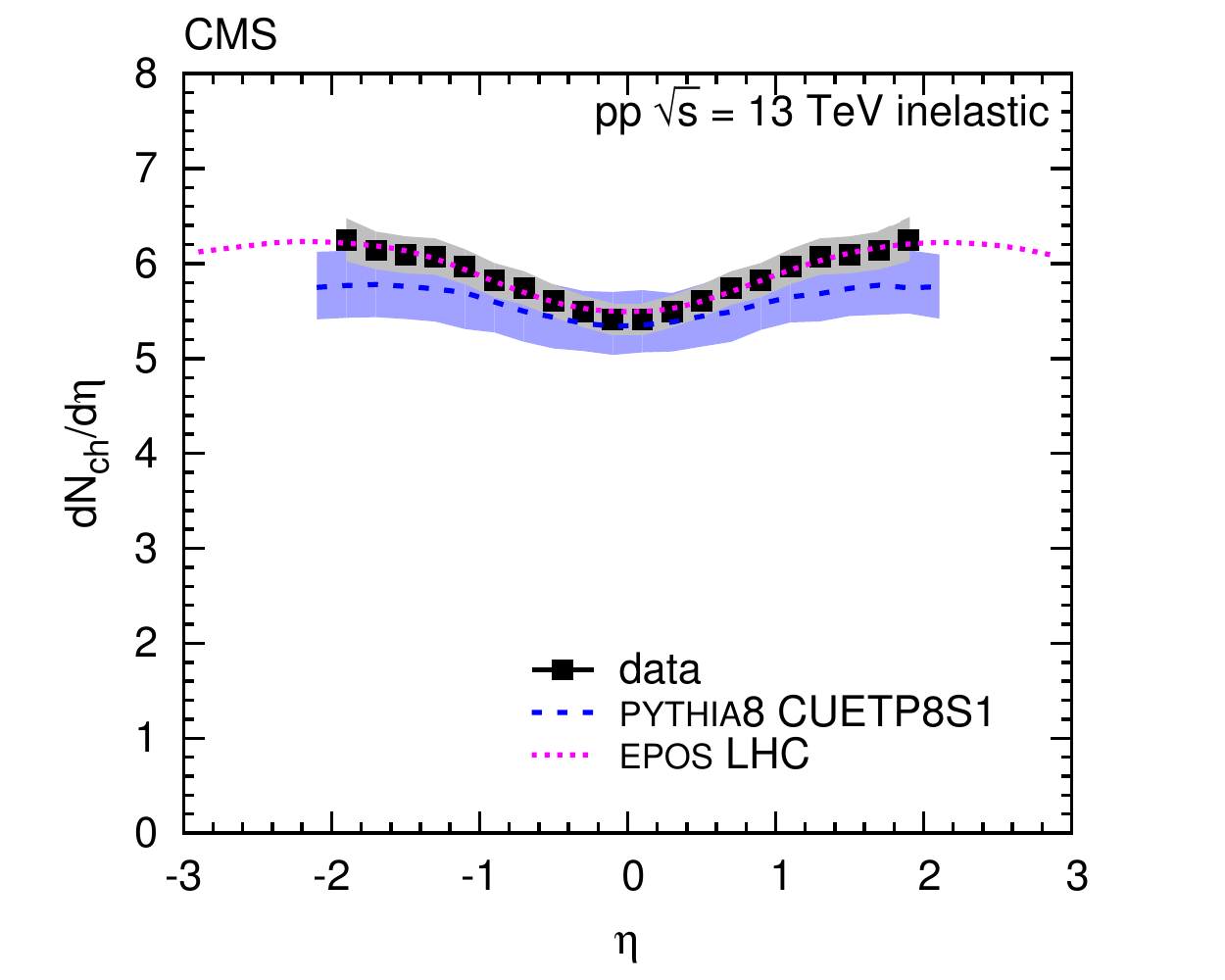}
  \includegraphics[width=0.49\textwidth]{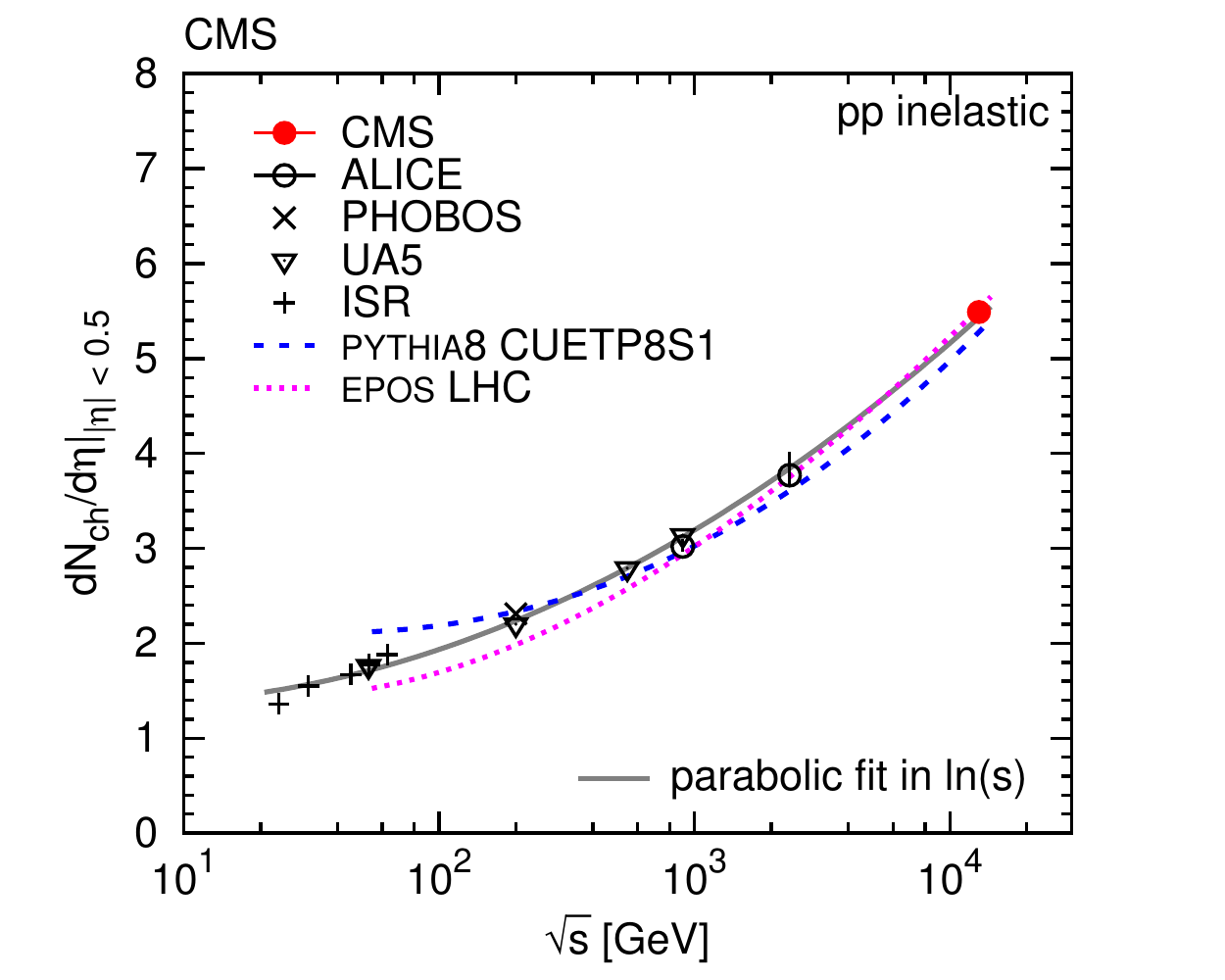}
 \caption{(\cmsLeft) Distributions of the pseudorapidity density of charged hadrons
in the region $\abs{\eta} < \meta$ in inelastic pp collisions at 13\TeV measured in
data (solid markers, combined track and tracklet results, symmetrized in
$\eta$), and predicted by the \Peight CUETP8S1 and the \EPOS event generators
(curves). The grey shaded area encompassing the data points indicates their
correlated systematic uncertainties. The blue band corresponds to the
envelope of the CUETP8S1 tune parametric uncertainties.
(\cmsRight) Center-of-mass energy dependence of $\dnchdeta|_{\abs{\eta} < 0.5}$
including ISR~\cite{Rossi:1974if,Thome:1977ky},
UA5~\cite{Alner:1986xu,Alner:1987wb}, PHOBOS~\cite{Back:2003hx}, and
ALICE~\cite{Aamodt:2010ft} data. The solid curve shows a second-order
polynomial in $\ln(s)$ fit to the data points, including the new result at
$\sqrt{s} = 13\TeV$. The dashed and dotted curves show the \Peight CUETP8S1 and
\EPOS predictions, respectively.}
\label{fig:dndeta_final}
\end{figure}

Pseudorapidity density distributions of charged hadrons in the region $\abs{\eta} <
\meta$ for inelastic pp events are shown in Fig.~\ref{fig:dndeta_final}~(\cmsLeft).
The data points and uncertainties are symmetrized in ${\pm}\eta$.
In the range $\abs{\eta} < 0.5$ the average pseudorapidity density is
$\dnchdeta|_{\abs{\eta} < 0.5} = \finaldat $.
While the predictions of both \Peight (with CUETP8S1 and CUETP8M1) and \EPOS
agree with the measured central value, the measured $\dnchdeta$ distribution in
the full $\eta$ range is better described by the latter. The uncertainty band
of \Peight corresponds to the envelope of the uncertainties of the tune
parameters of CUETP8S1; the \EPOS predictions have no uncertainty associated
with its parameter settings.

The center-of-mass energy dependence of \dnchdeta\ is shown in
Fig.~\ref{fig:dndeta_final}~(\cmsRight). For comparison with the $\sqrt{s} =
13$\TeV results presented in this Letter, inelastic pp measurements at lower
energies (ISR~\cite{Rossi:1974if,Thome:1977ky},
UA5~\cite{Alner:1986xu,Alner:1987wb}, PHOBOS~\cite{Back:2003hx}, and
ALICE~\cite{Aamodt:2010ft}) are also plotted.
The measured values are empirically fitted using a second-order polynomial in
$\ln(s)$ as $1.55 - 0.113 \ln(s) + 0.0168 \ln(s)^2$, where $s$ has the units
$\GeVns^2$, which provides a good description of the available data over the
full energy range.
The \Peight and \EPOS\ event generators globally reproduce the
collision-energy dependence of hadron production in inelastic pp collisions.

\section{Summary}

The pseudorapidity distribution of charged hadrons has been measured by the CMS
experiment, operated at zero magnetic field, at the LHC in proton-proton
collisions at $\sqrt{s} =13$\TeV.
Using two methods, based on hit pairs and straight-line tracks in the barrel
region of the CMS pixel detector, a charged hadron multiplicity at midrapidity,
$\dnchdeta|_{\abs{\eta} < 0.5} = \finaldat$, has been obtained for
inelastic pp events.
In the central region, the measured $\dnchdeta$ distribution is consistent with
predictions of the \Peight (with the CMS underlying event tunes CUETP8S1 and
CUETP8M1) and \EPOS\ (LHC tune) event generators, while those in a wider
$\eta$ range are better described by the latter. These results constitute the
first CMS measurement of hadron production at the new center-of-mass energy
frontier, and provide new constraints for the improvement of perturbative and
nonperturbative QCD aspects implemented in hadronic event generators.

\begin{acknowledgments}
We congratulate our colleagues in the CERN accelerator departments for the excellent performance of the LHC and thank the technical and administrative staffs at CERN and at other CMS institutes for their contributions to the success of the CMS effort. In addition, we gratefully acknowledge the computing centres and personnel of the Worldwide LHC Computing Grid for delivering so effectively the computing infrastructure essential to our analyses. Finally, we acknowledge the enduring support for the construction and operation of the LHC and the CMS detector provided by the following funding agencies: BMWFW and FWF (Austria); FNRS and FWO (Belgium); CNPq, CAPES, FAPERJ, and FAPESP (Brazil); MES (Bulgaria); CERN; CAS, MoST, and NSFC (China); COLCIENCIAS (Colombia); MSES and CSF (Croatia); RPF (Cyprus); MoER, ERC IUT and ERDF (Estonia); Academy of Finland, MEC, and HIP (Finland); CEA and CNRS/IN2P3 (France); BMBF, DFG, and HGF (Germany); GSRT (Greece); OTKA and NIH (Hungary); DAE and DST (India); IPM (Iran); SFI (Ireland); INFN (Italy); MSIP and NRF (Republic of Korea); LAS (Lithuania); MOE and UM (Malaysia); CINVESTAV, CONACYT, SEP, and UASLP-FAI (Mexico); MBIE (New Zealand); PAEC (Pakistan); MSHE and NSC (Poland); FCT (Portugal); JINR (Dubna); MON, RosAtom, RAS and RFBR (Russia); MESTD (Serbia); SEIDI and CPAN (Spain); Swiss Funding Agencies (Switzerland); MST (Taipei); ThEPCenter, IPST, STAR and NSTDA (Thailand); TUBITAK and TAEK (Turkey); NASU and SFFR (Ukraine); STFC (United Kingdom); DOE and NSF (USA). 

Individuals have received support from the Marie-Curie programme and the European Research Council and EPLANET (European Union); the Leventis Foundation; the A. P. Sloan Foundation; the Alexander von Humboldt Foundation; the Belgian Federal Science Policy Office; the Fonds pour la Formation \`a la Recherche dans l'Industrie et dans l'Agriculture (FRIA-Belgium); the Agentschap voor Innovatie door Wetenschap en Technologie (IWT-Belgium); the Ministry of Education, Youth and Sports (MEYS) of the Czech Republic; the Council of Science and Industrial Research, India; the HOMING PLUS programme of the Foundation for Polish Science, cofinanced from European Union, Regional Development Fund; the Compagnia di San Paolo (Torino); the Consorzio per la Fisica (Trieste); MIUR project 20108T4XTM (Italy); the Thalis and Aristeia programmes cofinanced by EU-ESF and the Greek NSRF; the National Priorities Research Program by Qatar National Research Fund; the Rachadapisek Sompot Fund for Postdoctoral Fellowship, Chulalongkorn University (Thailand); and the Welch Foundation.
\end{acknowledgments}

\bibliography{auto_generated}

\cleardoublepage \appendix\section{The CMS Collaboration \label{app:collab}}\begin{sloppypar}\hyphenpenalty=5000\widowpenalty=500\clubpenalty=5000\input{FSQ-15-001-authorlist.tex}\end{sloppypar}
\end{document}

%% file: FSQ-15-001-authorlist.tex
\textbf{Yerevan Physics Institute,  Yerevan,  Armenia}\\*[0pt]
V.~Khachatryan, A.M.~Sirunyan, A.~Tumasyan
\vskip\cmsinstskip
\textbf{Institut f\"{u}r Hochenergiephysik der OeAW,  Wien,  Austria}\\*[0pt]
W.~Adam, E.~Asilar, T.~Bergauer, J.~Brandstetter, E.~Brondolin, M.~Dragicevic, J.~Er\"{o}, M.~Flechl, M.~Friedl, R.~Fr\"{u}hwirth\cmsAuthorMark{1}, V.M.~Ghete, C.~Hartl, N.~H\"{o}rmann, J.~Hrubec, M.~Jeitler\cmsAuthorMark{1}, V.~Kn\"{u}nz, A.~K\"{o}nig, M.~Krammer\cmsAuthorMark{1}, I.~Kr\"{a}tschmer, D.~Liko, T.~Matsushita, I.~Mikulec, D.~Rabady\cmsAuthorMark{2}, B.~Rahbaran, H.~Rohringer, J.~Schieck\cmsAuthorMark{1}, R.~Sch\"{o}fbeck, J.~Strauss, W.~Treberer-Treberspurg, W.~Waltenberger, C.-E.~Wulz\cmsAuthorMark{1}
% ...
\vskip\cmsinstskip
% ...
1:~~Also at Vienna University of Technology, Vienna, Austria\\
2:~~Also at CERN, European Organization for Nuclear Research, Geneva, Switzerland\\
% ...
